\documentclass[9pt,twocolumn,twoside]{opticajnl}
\journal{opticajournal} 

\setboolean{shortarticle}{true}

\usepackage{lineno}

\title{The role of modes in nonlinear fiber optical computing}

\author[1]{Firdevs Yüce}
\author[2]{Bora Çarpınlıoğlu}
\author[2,*]{Uğur Teğin}

\affil[1]{Koç University Department of Mechanical Engineering, İstanbul, Türkiye}
\affil[2]{Koç University Department of Electrical and Electronics Engineering, İstanbul, Türkiye}

\affil[*]{utegin@ku.edu.tr}

\begin{abstract}
We  investigate the nonlinear propagation of light in graded-index multimode fiber, utilizing it as an optical computing unit, and quantify how it employs waveguide modes to process information. Using a time-dependent spatiotemporal propagation model with modal decomposition, we evaluate several benchmark regression and classification tasks and study the modal content of the generated speckles, which couples with a simple digital layer to perform optical computing. Analysis of modal entropy and energy-based mode counts reveals that effective computation is confined to a low-dimensional modal subspace, whose identity depends on the task and propagation regime. This also sets a trade-off between modal richness and nonlinear beam self-cleaning. These results establish modal statistics as practical design metrics for fiber-based optical computers.
\end{abstract}

\setboolean{displaycopyright}{false} 

\begin{document}

\maketitle

The exponential growth in computational demand for machine learning and artificial intelligence has exposed the fundamental limitations of traditional von Neumann digital architectures, particularly in terms of power dissipation and data transfer latency \cite{strubell2020energy}. The concept of optical computing is not new; foundational works established early on that the intrinsic properties of light—such as massive parallelism and passive linear operations—could theoretically offer energy-efficient information processing \cite{farhat1985optical,mcaulay1991optical,athale2016optical}. However, the mathematical framework for predetermined random and/or nonlinear projection is often referred to as reservoir computing and extreme learning machines. It was not until the exponential growth in computational demand for modern AI exposed the power dissipation and latency limitations of von Neumann architectures that these optical paradigms were revisited with urgency. This "optical renaissance" was marked by seminal demonstrations of photonic neural networks (PNNs) using free-space diffractive optics \cite{lin2018all,ccarpinliouglu2025genetically} and the subsequent development of integrated photonic circuits \cite{feldmann2021parallel,wetzstein2020inference}. While free-space and integrated platforms showed promise, they faced challenges regarding form factors and scaling limits \cite{alqadasi2022scaling}. 

More recently, attention has shifted from purely linear optical projections toward exploiting the computational potential of optical nonlinearities. Beyond linear mappings, a parallel line of work has explored how complex optical hardware itself can serve as a nonlinear computational substrate, where learning emerges from the underlying physical dynamics rather than being explicitly programmed. In this direction, studies of multimode fiber (MMF) propagation revealed rich spatial modal interactions and spatiotemporal dynamics that can be harnessed for computation \cite{tegin2021scalable}. Closely related efforts broadened this perspective through the concept of physical neural networks, in which controllable nonlinear physical systems—spanning optics, mechanics, and electronics—are directly trained using hybrid in-situ and in-silico backpropagation schemes \cite{wright2022deep}. These approaches elevate nonlinear wave interactions and physical coupling to the level of computational primitives, effectively shifting part of the learning process from electronic processors into the physical domain itself. Consequently, state-of-the-art research now focuses on programming the input wavefront to tailor these interactions for specific tasks \cite{oguz2024programming}, as well as exploiting chaotic propagation to reduce the power requirements of nonlinear optical computing \cite{kesgin2025photonic}.

Recent studies further demonstrate that spatiotemporal optical dynamics can directly emulate nonlinear activation functions in hardware \cite{yildirim2023nonlinear, redding2024fiber}. However, this progress exposes a fundamental paradox, while sufficiently high optical power is required to induce nonlinear interactions, excessive power can trigger beam self-cleaning, collapse modal diversity, and ultimately degrade computational performance \cite{yildirim2024nonlinear,  hary2025principles, manuylovich2025optical, muda2025optical}. As a result, the central challenge in fiber-based optical computing lies in achieving a careful balance between nonlinear strength and modal diversity, enabling robust, scalable, and power-efficient analog processors.

Here, we numerically investigate a nonlinear graded-index MMF as an optical computing unit and quantify how it uses its guided modes to process information. Using a time-dependent propagation model together with modal decomposition, we evaluate the regression and classification performance of the nonlinear fiber optical computing and show that nonlinear multimode propagation transforms the inputs into speckle-like features that a simple labeling readout can exploit. By analyzing modal entropy and energy-based mode counts, we find that the effective computation is confined to a low-dimensional modal subspace whose identity depends on the task and dataset. We further identify a trade-off between modal richness and nonlinear beam self-cleaning, stronger cumulative nonlinearity and longer propagation reconcentrate power into the lowest-order modes. These results establish modal statistics as practical figures of merit for designing and operating fiber-based optical computers in task-dependent regimes.

We model nonlinear propagation in an ideal graded-index MMF by solving a (3+1)D time-dependent beam propagation equation for the slowly varying envelope $A(x,y,t)$ at a carrier wavelength $\lambda_c = 1030~\mathrm{nm}$. The evolution along $z$ is described by
\begin{equation}
\partial_z A = 
\left[
\frac{i}{2k_0}\nabla_{\perp}^2
- i \frac{k_0 \delta}{R^2}(x^2+y^2)
- i\frac{\beta_2}{2}\partial_t^2
+ \frac{\beta_3}{6}\partial_t^3
+ i\gamma |A|^2
\right] A ,
\end{equation}
where $k_0 = 2\pi n_0/\lambda_c$ is the wavenumber in the cladding, $n_0 = 1.44$ is the refractive index, $R = 25~\mu\mathrm{m}$ is the core radius, and $\delta = 0.01$ sets the index contrast of the fiber. Group-velocity dispersion and third-order dispersion are included through $\beta_2 = 24.8\times 10^{-27}~\mathrm{s^2/m}$ and $\beta_3 = 23.3\times 10^{-42}~\mathrm{s^3/m}$, respectively. Kerr nonlinearity is modeled by $\gamma = 2\pi n_2/\lambda_c$ with $n_2 = 3.2\times 10^{-20}~\mathrm{m^2/W}$.

The temporal window is $T = 20~\mathrm{ps}$, sampled with $N_t = 2^{11}$ points, and the transverse window is $W = 54.1442~\mu\mathrm{m}$ in both $x$ and $y$, sampled with $N_x = N_y = 64$ points. To ensure reasonable simulation times while maintaining significant nonlinearity, we set the fiber length to $L_\mathrm{short} = 5.55~\mathrm{mm}$ and the peak power of the input beam to $P_\mathrm{peak} = 3~\mathrm{MW}$. Information is encoded as phase-only modulation of a Gaussian input beam with transverse FWHM $25~\mu\mathrm{m}$ and temporal FWHM $1~\mathrm{ps}$ at $z=0$. All simulations for the different datasets use the same fiber and numerical parameters.

To analyze the role of guided modes, we perform a modal decomposition of the fields at different propagation distances. For the ideal graded-index MMF, we compute an orthonormal set of guided spatial LP modes $u_m(x,y)$ at the carrier wavelength and find that the test fiber supports 120 modes. At the end of the fiber we compute the time-integrated complex field $E_n(x,y)$ for each sample $n$ in the dataset, and decompose it onto the fiber modes to obtain modal amplitudes \cite{rothe2020deep}. 
\begin{equation}
c_{m,n} = \iint E_n(x,y)\, u_m^*(x,y)\,\mathrm{d}x\,\mathrm{d}y ,
\end{equation}
from which we define the normalized modal power fractions
\begin{equation}
p_{m,n}(z) = \frac{|c_{m,n}(z)|^2}{\sum_{m'} |c_{m',n}(z)|^2}.
\end{equation}
Stacking $p_{m,n}$ over samples $n$ yields the sample--mode power matrices used in the figures. For each sample we define the modal Shannon entropy \cite{paur2016achieving}.
\begin{equation}
H_n(z) = -\sum_m p_{m,n}(z)\,\log_2 p_{m,n}(z),
\end{equation}
and the corresponding effective number of modes
\begin{equation}
N_{\mathrm{eff},n}(z) = 2^{H_n(z)}.
\end{equation}
The distributions of $N_{\mathrm{eff}}$ across samples and datasets quantify how many fiber modes are effectively used by the nonlinear optical computer at a given propagation length.

\begin{figure}[t]
\centering
\includegraphics[width=\linewidth]{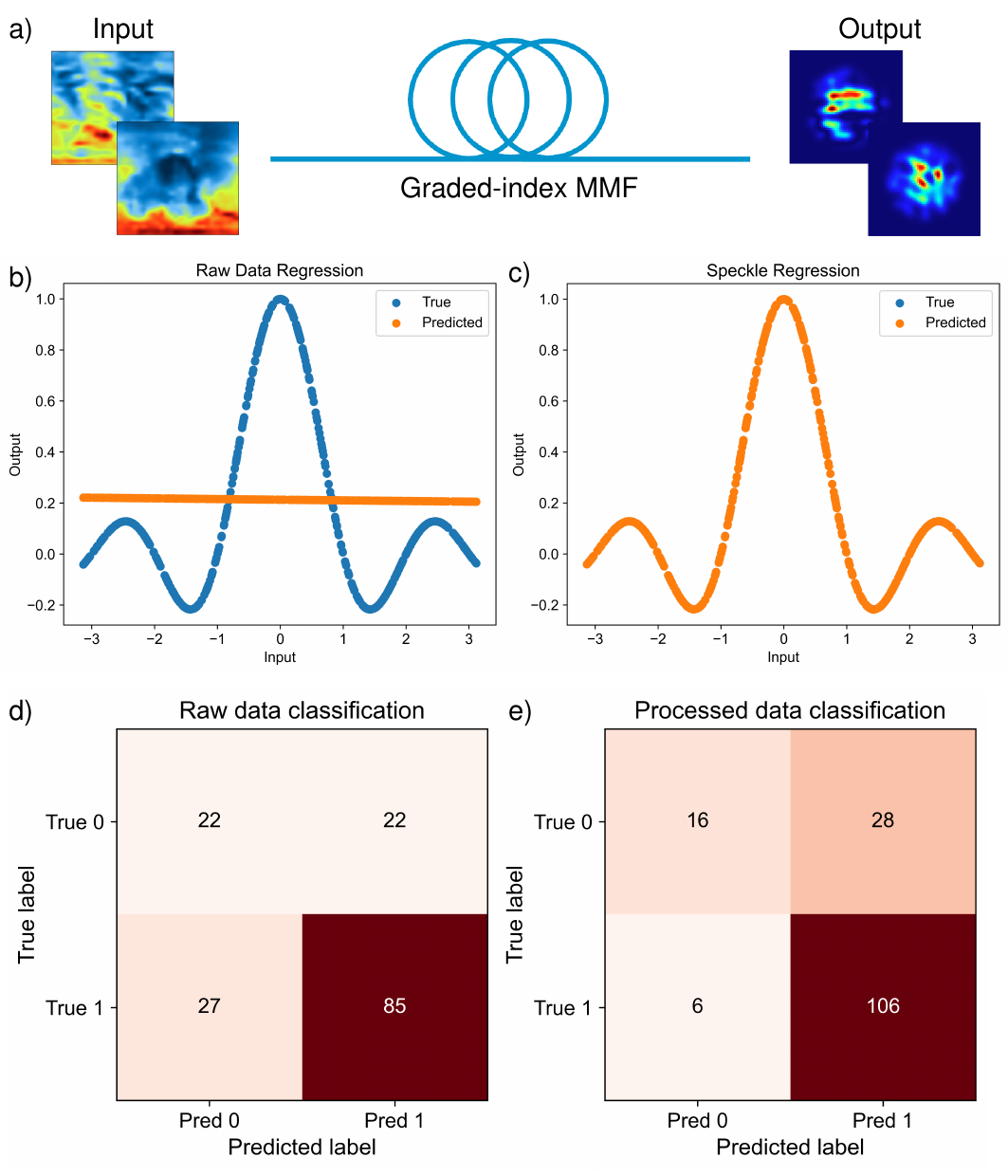}
\caption{Schematic of the fiber optical computing unit and performance on regression and classification tasks. (a) System architecture. (b) and (c) Sinc regression task before and after optical computing. (d) and (e)  BreastMNIST classification task before and after optical computing.}
\label{fig:system_performance}
\end{figure}

We evaluate the performance of the nonlinear fiber optical computing unit on two distinct tasks. For Sinc regression, applying a linear regressor to the output intensity patterns yields a coefficient of determination ($R^2$) of $99.9\%$ on test data, whereas training directly on raw scalar inputs results in a negligible score of $0.013\%$. For the BreastMNIST classification task, the optical unit achieves a test accuracy of $78.21\%$, compared to $68.59\%$ when the classifier is trained on the raw dataset. In both cases, nonlinear multimode propagation transforms the inputs into high-dimensional speckle features that a simple linear model can exploit, providing a substantial gain over the purely electronic baseline with the same readout architecture.

\begin{figure}[t]
\centering
\includegraphics[width=\linewidth]{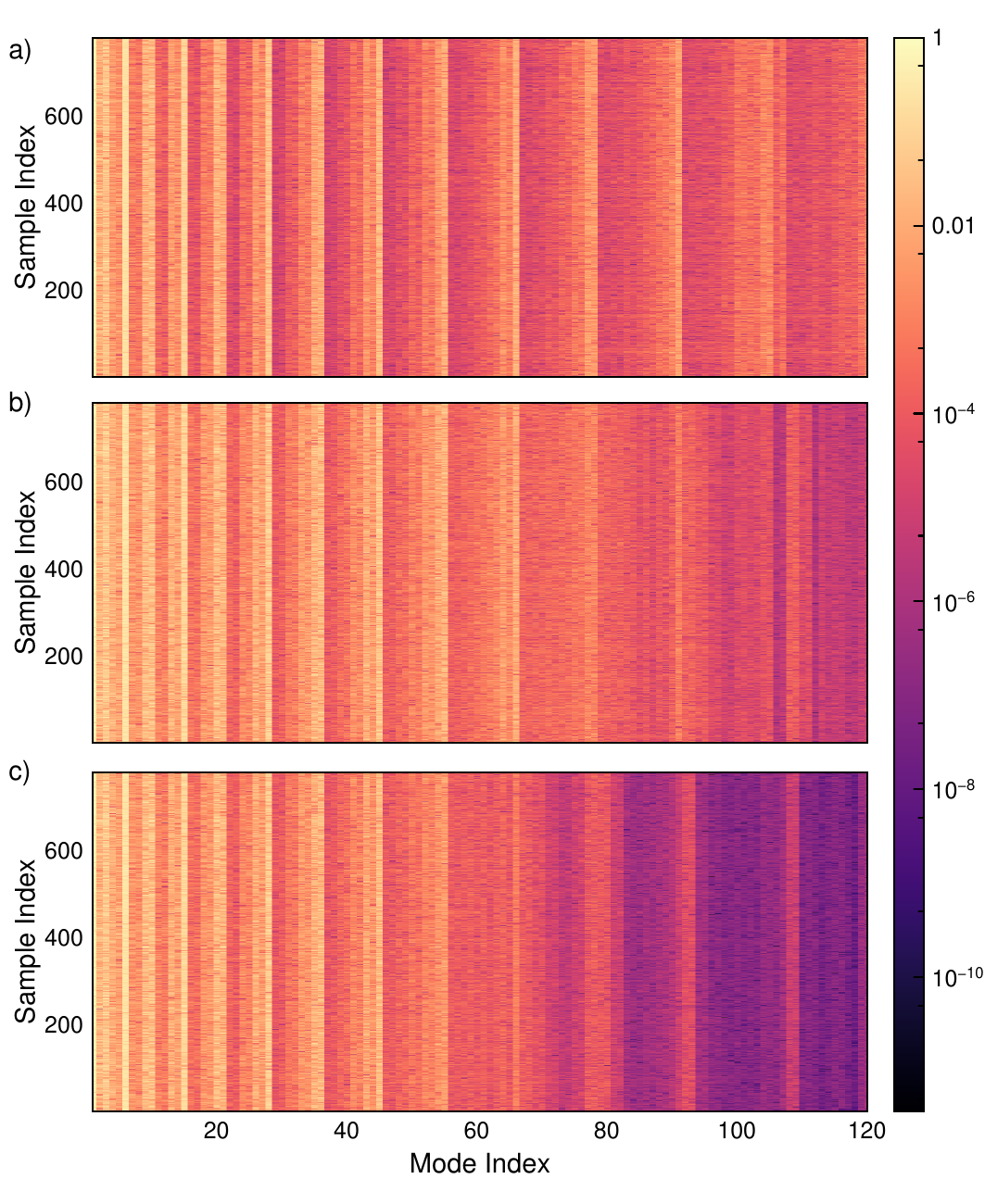}
\caption{ Modal power fraction matrices for the BreastMNIST task at different propagation regimes. Each panel shows the normalized modal power fractions $p_{m,n}$  on a logarithmic scale for (a) near the input (after a single propagation step, 555 \textmu m), (b) 5.55 mm with 3 MW peak powers, and (c) a longer propagation distance (2.75 cm) with reduced peak power (600 kW).}
\label{fig:modal_matrices}
\end{figure}

To analyze how nonlinear propagation redistributes power among the guided modes, we decompose the speckle patterns obtained from the BreastMNIST dataset at various propagation distances. We compute the normalized modal power fraction $p_{m,n}(z)$ for each input sample $n$ and guided mode $m$, stacking these values into matrices to visualize mode occupancy.

Fig.~\ref{fig:modal_matrices}(a) displays the modal power fractions near the input (after a single step $\Delta z$). Even at this early stage, power is confined to a small subset of low-order modes, visible as bright vertical bands, while high-order modes remain several orders of magnitude weaker. This confirms that the input encoding populates only a limited subset of the modal basis.

At the computing length ($L_\mathrm{short}=5.55~\mathrm{mm}$) with a peak power of $P_\mathrm{peak}=3~\mathrm{MW}$, nonlinear intermodal coupling becomes evident (Fig.~\ref{fig:modal_matrices}(b)). Power spreads from the fundamental mode into a cluster of neighboring low-order modes, which become clearly visible on the logarithmic scale, while many higher-order modes remain weakly excited. Thus, nonlinear propagation activates specific guided modes beyond those initially excited by the input coupling, but still within a restricted subset of the full modal basis.

To probe a stronger propagation regime, we increase the fiber length five-fold while proportionally decreasing the peak power (Fig.~\ref{fig:modal_matrices}(c)). In this configuration, the power distribution trends shift, instead of continuing to spread, energy reconcentrates in the lowest-order modes, while neighboring modes appear darker on the logarithmic scale. This behavior is consistent with the onset of nonlinear beam self-cleaning, where the graded-index fiber guides energy back into the fundamental modes over longer distances.

\begin{figure}[t]
\centering
\includegraphics[width=\linewidth]{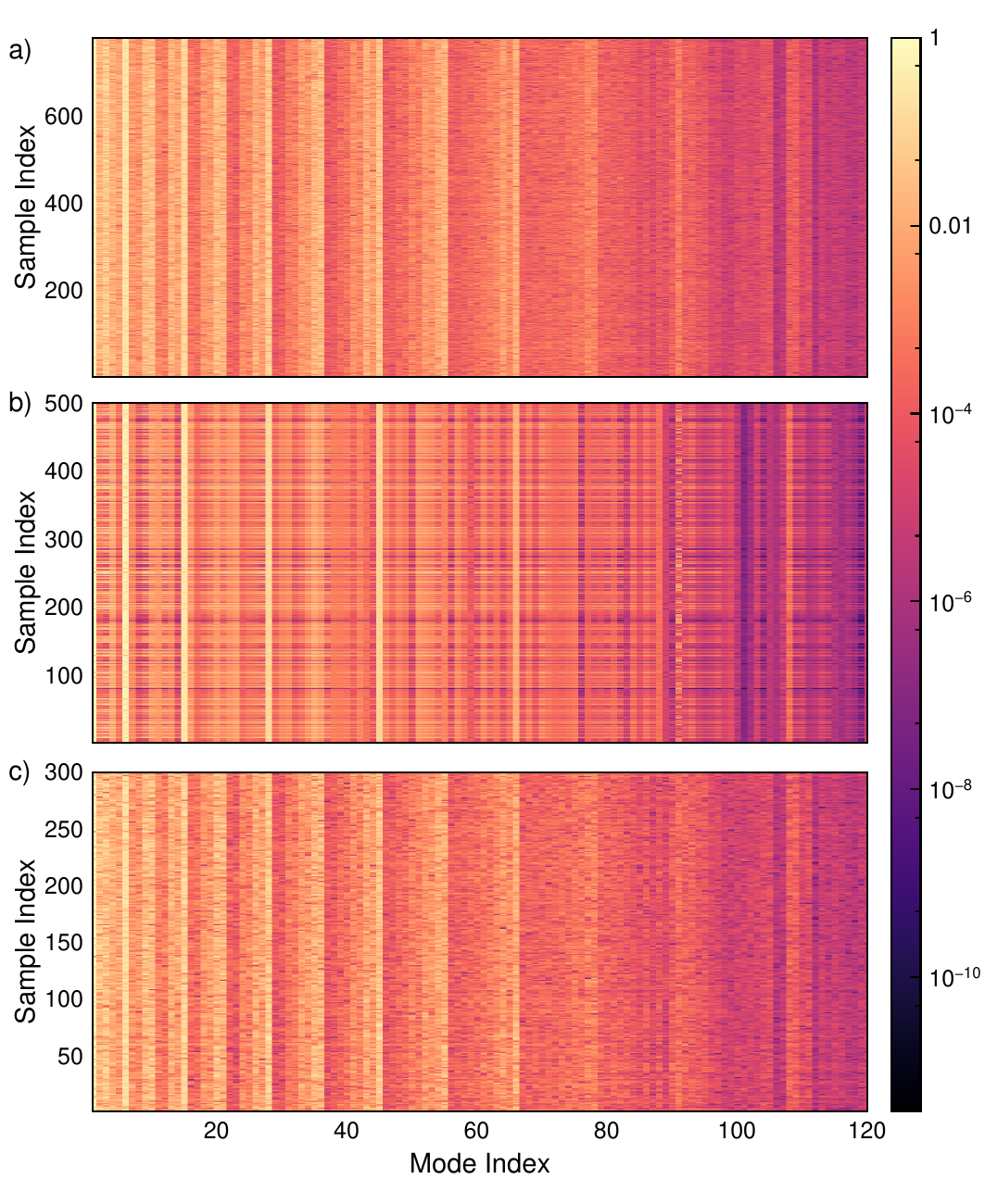}
\caption{Dataset-dependent modal power fraction matrices. Normalized modal power fractions $p_{m,n}$ on a logarithmic scale for (a) BreastMNIST, (b) Sinc, and (c) small-CIFAR (airplane, cat, and dog classes) after 5.55 mm fiber length with 3 MW peak powers, enabling direct comparison of modal occupancy across tasks.}
\label{fig:dataset_matrices}
\end{figure}

We next examine how different datasets populate the guided modes after nonlinear propagation under identical parameters. We compute the mode power fraction matrices $p_{m,n}$ for the Sinc, BreastMNIST, and small-CIFAR (airplane, cat, and dog classes) datasets. Fig.~\ref{fig:dataset_matrices} presents these matrices on a logarithmic scale for direct comparison across tasks.

Compared to BreastMNIST (Fig.~\ref{fig:dataset_matrices}(a)), the Sinc dataset (Fig.~\ref{fig:dataset_matrices}(b)) exhibits a more concentrated modal distribution, with power confined to a tight cluster of modes near the fundamental. This reflects the effectively one-dimensional nature of the Sinc task, scalar inputs encoded via a single random mask primarily populate a few low-order modes, resulting in limited nonlinear excitation of higher-order states.

In contrast, the small-CIFAR dataset (Fig.~\ref{fig:dataset_matrices}(c)) engages a significantly broader set of modes. The matrix reveals extended regions of high intensity across the mode index, with a larger number of modes reaching appreciable power levels on the logarithmic scale. The complex, natural structures in CIFAR images therefore excite a richer subset of the fiber's modal basis, leading to a wider redistribution of power than in either the Sinc or BreastMNIST cases.

\begin{figure}[t]
\centering
\includegraphics[width=\linewidth]{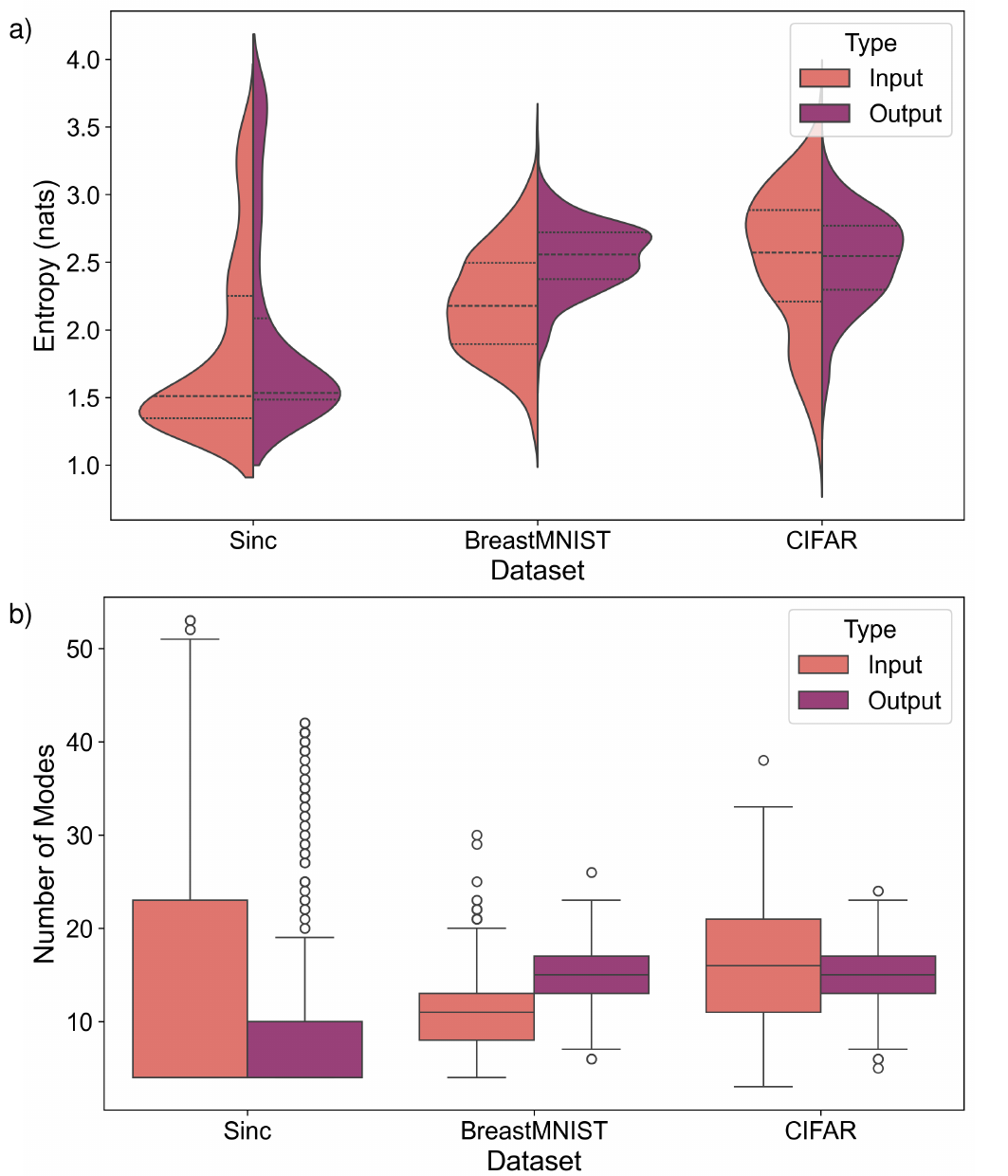}
\caption{Statistics of modal usage. (a) Distributions of modal Shannon entropy across samples and datasets. (b) Distributions of the number of modes required to capture $90\%$ of the total energy, quantifying how many modes are effectively used by the nonlinear optical computer.}
\label{fig:false-color}
\end{figure}

Our results show that, although the multimode fiber supports many guided modes, the effective computation takes place in a low-dimensional modal subspace. The modal power matrices in Figs.~\ref{fig:modal_matrices} and \ref{fig:dataset_matrices} already indicate that only a restricted set of modes carries most of the energy, with many higher-order modes remaining several orders of magnitude weaker on the logarithmic scale. Fig.~\ref{fig:false-color}(a) quantifies this observation in terms of modal entropy: for all datasets and propagation conditions, the distributions remain at comparatively low values, indicating that the output fields occupy only a small fraction of the available modal degrees of freedom. The corresponding distributions of the number of modes required to capture $90\%$ of the total energy in Fig.~\ref{fig:false-color}(b) reinforce this picture.

Within this low-dimensional manifold, the observed specific modal subspace is task dependent as illustrated via the comparison between Sinc, BreastMNIST, and small-CIFAR datasets in Fig.~\ref{fig:dataset_matrices}. The Sinc task excites a very narrow band of low-order modes and exhibits the lowest modal entropy and smallest $90\%$-energy mode count in Fig.~\ref{fig:false-color}. BreastMNIST, with higher input dimensionality than Sinc inputs, populates a broader set of modes and shifts the entropy and mode-count distributions to higher values. The small-CIFAR dataset engages the richest modal structure, with extended regions of non-negligible power across the mode index and correspondingly higher modal entropy and larger $90\%$-energy mode counts. Tasks with more complex input structure therefore require the optical system to exploit a richer subset of the modal basis to generate linearly separable features at the output; the fiber does not use a fixed “universal” modal feature map, but adapts the relevant subspace to the dataset.

The propagation regime and nonlinearity further modulate how much of the modal basis is actually used. At the shorter computing length with higher peak power, the nonlinear Kerr effect and intermodal coupling broaden the initially populated set of modes (Fig.~\ref{fig:modal_matrices}(b)), increasing modal entropy compared to the near-input distribution in Fig.~\ref{fig:modal_matrices}(a). When we move to longer propagation with reduced peak power, the system instead approaches a nonlinear beam-cleaning regime (Fig.~\ref{fig:modal_matrices}(c)), in which energy reconcentrates in the lowest-order modes. Consistently, a reduction in modal entropy and in the number of modes required to reach $90\%$ energy in this regime. Strong cumulative nonlinearity and long propagation thus trade modal richness for beam self-cleaning, effectively collapsing the computation onto an even smaller modal subspace. It creates many-to-one mapping similar to \cite{muda2025optical} and decreases optical computing performance.

These observations highlight a design trade-off for fiber-based optical computing. Richer modal occupation enlarges the feature space and can increase expressivity, particularly for high-dimensional inputs such as natural images, whereas operating too deep in the nonlinear, long-propagation regime drives the system toward beam cleaning, which reduces modal entropy and the diversity of features accessible to the linear readout. Viewing the multimode fiber as a physical feature map suggests using modal statistics such as entropy and the $90\%$-energy mode count as diagnostics for choosing fiber length, input power, and coupling conditions, and motivates a co-design of optics and electronics that is aware of these task-dependent modal manifolds.

We have numerically investigated a nonlinear graded-index MMF as an optical computing unit, focusing on how it uses its guided modes to implement task-dependent feature maps. Modal entropy and energy-based mode counts show that the computation effectively lives in a small modal subspace, even though the fiber supports many guided modes. The identity of this subspace depends on the task: simple, low-dimensional inputs largely occupy a narrow band of low-order modes, whereas more complex image data engage a richer subset of the modal basis. The propagation regime and nonlinearity then control a trade-off between modal richness and nonlinear beam self-cleaning, with intermediate regimes providing diverse speckle features and stronger cumulative nonlinearity reconcentrating power into the lowest-order modes. These results position modal statistics as practical figures of merit for designing fiber-based optical computers and suggest co-design strategies for nonlinear multimode propagation, input encoding, and electronic readout that deliberately target task-specific modal manifolds.
\begin{backmatter}

\bmsection{Funding} Türkiye Bilimsel ve Teknolojik Araştırma Kurumu (122C150).

\bmsection{Disclosures} The authors declare no conflicts of interest.

\bmsection{Data availability} Data underlying the results presented in this paper are not publicly available at this time but may be obtained from the authors upon reasonable request.

\end{backmatter}

\bibliography{sample}

@inproceedings{strubell2020energy,
  title={Energy and policy considerations for modern deep learning research},
  author={Strubell, Emma and Ganesh, Ananya and McCallum, Andrew},
  booktitle={Proceedings of the AAAI Conference on Artificial Intelligence},
  volume={34},
  number={09},
  pages={13693--13696},
  year={2020},
  note={}
}

@article{athale2016optical,
  title={Optical computing: past and future},
  author={Athale, Ravi and Psaltis, Demetri},
  journal={Optics and Photonics News},
  volume={27},
  number={6},
  pages={32--39},
  year={2016},
  publisher={OSA}
}

@article{farhat1985optical,
  title={Optical implementation of the Hopfield model},
  author={Farhat, Nabil H and Psaltis, Demetri and Prata, Aluizio and Paek, Eung},
  journal={Applied optics},
  volume={24},
  number={10},
  pages={1469--1475},
  year={1985},
  publisher={Optical Society of America}
}

@article{wetzstein2020inference,
  title={Inference in artificial intelligence with deep optics and photonics},
  author={Wetzstein, Gordon and Ozcan, Aydogan and Gigan, Sylvain and Fan, Shanhui and Englund, Dirk and Solja{\v{c}}i{\'c}, Marin and Denz, Cornelia and Miller, David AB and Psaltis, Demetri},
  journal={Nature},
  volume={588},
  number={7836},
  pages={39--47},
  year={2020},
  publisher={Nature Publishing Group UK London}
}

@article{tegin2021scalable,
  title={Scalable optical learning operator},
  author={Te\u{g}in, U\u{g}ur and Yildirim, Mustafa and O\u{g}uz, Ilker and Moser, Christophe and Psaltis, Demetri},
  journal={Nature Computational Science},
  volume={1},
  number={8},
  pages={542--549},
  year={2021},
  note={}
}

@article{kesgin2025photonic,
  title={Photonic neural networks at the edge of spatiotemporal chaos in multimode fibers},
  author={Kesgin, Bahad{\i}r Utku and Te{\u{g}}in, U{\u{g}}ur},
  journal={Nanophotonics},
  pages = {2723--2732},
  volume = {14},
  number = {16},
  year={2025},
  publisher={De Gruyter}
}

@article{ccarpinliouglu2025genetically,
  title={Genetically programmable optical random neural networks},
  author={{\c{C}}arp{\i}nl{\i}o{\u{g}}lu, Bora and Te{\u{g}}in, U{\u{g}}ur},
  journal={Communications Physics},
  volume={8},
  number={1},
  pages={349},
  year={2025},
  publisher={Nature Publishing Group UK London}
}

@article{oguz2024programming,
  title={Programming nonlinear propagation for efficient optical learning machines},
  author={Oguz, Ilker and Hsieh, Jih-Liang and Dinc, Niyazi Ulas and Teğin, Uğur and Yildirim, Mustafa and Gigli, Carlo and Moser, Christophe and Psaltis, Demetri},
  journal={Advanced Photonics},
  volume={6},
  number={1},
  pages={016002},
  year={2024},
  note={}
}

@book{mcaulay1991optical,
  title={Optical Computer Architectures: The Application of Optical Concepts to Next Generation Computers},
  author={McAulay, Alastair D},
  year={1991},
  publisher={Wiley},
  note={}
}

@article{lin2018all,
  title={All-optical machine learning using diffractive deep neural networks},
  author={Lin, Xing and Rivenson, Yair and Yardimci, Nezih T and Veli, Muhammed and Luo, Yi and Jarrahi, Mona and Ozcan, Aydogan},
  journal={Science},
  volume={361},
  number={6406},
  pages={1004--1008},
  year={2018},
  note={}
}

@article{yildirim2024nonlinear,
  title={Nonlinear processing with linear optics},
  author={Yildirim, Mustafa and Dinc, Niyazi Ulas and Oguz, Ilker and Psaltis, Demetri and Moser, Christophe},
  journal={Nature Photonics},
  volume={18},
  pages={1076--1082},
  year={2024},
  note={}
}

@article{feldmann2021parallel,
  title={Parallel convolutional processing using an integrated photonic tensor core},
  author={Feldmann, J and Youngblood, N and Karpov, M and Gehring, H and Li, X and Stappers, M and Le Gallo, M and Fu, X and Lukashchuk, A and Raja, A and others},
  journal={Nature},
  volume={589},
  number={7840},
  pages={52--58},
  year={2021},
  note={}
}

@article{alqadasi2022scaling,
  title={Scaling up silicon photonic-based accelerators: challenges and opportunities},
  author={Al-Qadasi, M A and others},
  journal={APL Photonics},
  volume={7},
  number={2},
  pages={020902},
  year={2022},
  note={}
}

@article{wright2022deep,
  title={Deep physical neural networks trained with backpropagation},
  author={Wright, Logan G and Onodera, Tatsuhiro and Stein, Martin M and Wang, Tianyu and Schachter, Darren T and Hu, Zoey and McMahon, Peter L},
  journal={Nature},
  volume={601},
  number={7894},
  pages={549--555},
  year={2022},
  note={}
}

@article{redding2024fiber,
  title={Fiber optic computing using distributed feedback},
  author={Redding, Brandon and Murray, Joseph B and Hart, Joseph D and Zhu, Zheyuan and Pang, Shuo S and Sarma, Raktim},
  journal={Communications Physics},
  volume={7},
  pages={75},
  year={2024},
  note={}
}

@article{yildirim2023nonlinear,
  title={Nonlinear optical feature generator for machine learning},
  author={Yildirim, M and others},
  journal={APL Photonics},
  volume={8},
  number={10},
  pages={106104},
  year={2023},
  note={}
}

@article{muda2025optical,
  title={Optical computing with supercontinuum generation in photonic crystal fibers},
  author={Muda, Azka Maula Iskandar and Te\u{g}in, U\u{g}ur},
  journal={Optics Express},
  volume={33},
  number={4},
  pages={7852},
  year={2025},
  note={}
}

@article{hary2025principles,
  title={Principles and metrics of extreme learning machines using a highly nonlinear fiber},
  author={Hary, Mathilde and Brunner, Daniel and Leybov, Lev and Ryczkowski, Piotr and Dudley, John M and Genty, Goery},
  journal={Nanophotonics},
  pages = {2733--2748},
  volume = {14},
  number = {16},  year={2025},
  publisher={De Gruyter}
}

@article{manuylovich2025optical,
  title={Optical neuromorphic computing via temporal up-sampling and trainable encoding on a telecom device platform},
  author={Manuylovich, Egor and Stoliarov, Dmitrii and Saad, David and Turitsyn, Sergei K},
  journal={Nanophotonics},
  number={16},
  pages={2761--2778},
  year={2025},
  publisher={De Gruyter}
}

@article{rothe2020deep,
  title={Deep learning for computational mode decomposition in optical fibers},
  author={Rothe, Stefan and Zhang, Qian and Koukourakis, Nektarios and Czarske, Juergen W},
  journal={Applied Sciences},
  volume={10},
  number={4},
  pages={1367},
  year={2020},
  publisher={MDPI}
}

@article{paur2016achieving,
  title={Achieving the ultimate optical resolution},
  author={Paur, Martin and Stoklasa, Bohumil and Hradil, Zdenek and Sanchez-Soto, Luis L and Rehacek, Jaroslav},
  journal={Optica},
  volume={3},
  number={10},
  pages={1144--1147},
  year={2016},
  publisher={Optical Society of America}
}

\bibliographyfullrefs{sample}


\ifthenelse{\equal{\journalref}{aop}}{%
\section*{Author Biographies}
\begingroup
\setlength\intextsep{0pt}
\begin{minipage}[t][6.3cm][t]{1.0\textwidth} 
  \begin{wrapfigure}{L}{0.25\textwidth}
    \includegraphics[width=0.25\textwidth]{john_smith.eps}
  \end{wrapfigure}
  \noindent
  {\bfseries John Smith} received his BSc (Mathematics) in 2000 from The University of Maryland. His research interests include lasers and optics.
\end{minipage}
\begin{minipage}{1.0\textwidth}
  \begin{wrapfigure}{L}{0.25\textwidth}
    \includegraphics[width=0.25\textwidth]{alice_smith.eps}
  \end{wrapfigure}
  \noindent
  {\bfseries Alice Smith} also received her BSc (Mathematics) in 2000 from The University of Maryland. Her research interests also include lasers and optics.
\end{minipage}
\endgroup
}{}

\end{document}